\documentclass[fleqn,usenatbib]{mnras}
\usepackage{newtxtext,newtxmath}
\usepackage[T1]{fontenc}

\DeclareRobustCommand{\VAN}[3]{#2}
\let\VANthebibliography\thebibliography
\def\thebibliography{\DeclareRobustCommand{\VAN}[3]{##3}\VANthebibliography}

\usepackage{graphicx}	% Including figure files
\usepackage{amsmath}	% Advanced maths commands
\title[Searching for Corannulene with ALMA: Observations of the Red Rectangle Nebula]{Searching for Corannulene with ALMA: Observations of the Red Rectangle Nebula}

\author[Koo et al.]{Elise Koo$^{1}$, Alessandra Candian$^{1}$\thanks{a.candian@uva.nl}, Michiel Hogerheijde$^{1,2}$, Lizette Guzman-Ramirez$^{3}$, Javier Alcolea$^{4}$,
\newauthor Valentin Bujarrabal$^{4}$, Jan Cami$^{5,6,7}$, Pierre Cox$^{8}$, Peter Sarre$^{9}$ \\
$^{1}$University of Amsterdam, Anton Pannekoek Institute, Science Park 904, 1098XH, Netherlands\\
$^{2}$Leiden Observatory, Leiden University, PO Box 9513, 2300 RA Leiden, The Netherlands\\
$^{3}$Eindhoven University of Technology, Eindhoven 5600 MB, The Netherlands\\
$^{4}$Observatorio Astronómico Nacional (OAN-IGN), Ap. 112, 28803 Alcalá de Henares, Spain\\
$^{5}$Department of Physics \& Astronomy, The University of Western
Ontario, London ON N6A 3K7, Canada\\
$^{6}$Institute for Earth and Space Exploration, The University of Western
Ontario, London ON N6A 3K7, Canada\\
$^{7}$Carl Sagan Center, SETI Institute, 339 Bernardo Avenue, Suite 200,
Mountain View, CA 94043, USA\\
$^{8}$CNRS and Sorbonne Université, UMR 7095, Institut
d’Astrophysique de Paris, 98bis boulevard Arago, 75014 Paris,
France\\
$^{9}$School of Chemistry, The University of Nottingham, University
Park, Nottingham NG7 2RD, UK
}

\date{Accepted XXX. Received YYY; in original form ZZZ}

\pubyear{\the\year{}}

\begin{document}
\label{firstpage}
\pagerange{\pageref{firstpage}--\pageref{lastpage}}
\maketitle

% Abstract of the paper
\begin{abstract}
Polycyclic Aromatic Hydrocarbons (PAHs) are organic molecules responsible for the Aromatic Infrared Bands (AIBs), observed across a multitude of astrophysical environments. Despite their ubiquity, the precise formation mechanisms of PAHs remain unclear. One of the possible way for PAHs to form is in the outflows of evolved stars, such as HD 44179, which produces the Red Rectangle nebula—a known emitter of AIBs. However, no specific PAH molecules have been detected in such environments, complicating the understanding of PAH formation and evolution.
This study aimed to detect the PAH molecule corannulene (C$_{20}$H$_{10}$), a viable candidate for radio detection due to its large dipole moment of 2.07 D. We analyzed high-resolution band 4 ALMA observations of the Red Rectangle nebula, collected over almost 9 hrs. Although corannulene emission was not detected, we estimated a firm upper limit on its abundance compared to hydrogen (5 $\times$ 10$^{-13}$) and we discuss the lack of detection in the context of our current understanding of PAH formation and destruction mechanisms. Additionally, we report tentative detection of signals at 139.612 GHz, 139.617 GHz, and 139.621 GHz, potentially originating from cyclopropenyledine (c-C$_3$H$_2$) and the 140 GHz H$_2$O maser. 
\end{abstract}

% Select between one and six entries from the list of approved keywords.
% Don't make up new ones.
\begin{keywords}
Astrochemistry -- stars: individual: Red Rectangle -- ISM: molecules -- submillimetre: ISM 
\end{keywords}

%%%%%%%%%%%%%%%%%%%%%%%%%%%%%%%%%%%%%%%%%%%%%%%%%%

%%%%%%%%%%%%%%%%% BODY OF PAPER %%%%%%%%%%%%%%%%%%

\section{Introduction}
Polycyclic Aromatic Hydrocarbons (PAHs) are a class of complex organic molecules originally proposed by \cite{Leger} and \cite{Allamandola1984} as being responsible for the Aromatic Infrared Bands (AIBs). These emission features have been observed in many interstellar objects. For example, they are detected in the interstellar medium (ISM) of galaxies \citep{Draine2007, Lin25},
and play a critical role in the cooling and heating of the interstellar gas \citep{Tielens2008}.  Their sensitivity to the physical conditions of the local environment makes them valuable probes of UV-irradiated surfaces of molecular clouds \citep{Chown2024} and photospheres of protoplanetary disks \citep{Lange2023}. Due to the strength and widespread detection of the AIBs, it is estimated that up to 10\%  of all the carbon in the Universe is contained in PAH molecules with size larger than roughly 50 carbon atoms.  In addition, they have been widely studied experimentally as potential carriers of the Diffuse Interstellar Bands (DIBs), a set of absorption features observed in stellar spectra \citep{Sarre2006, Geballe2016}. PAHs are investigated also in the context of the Solar nebula. Not only they are detected in carbonaceous chondrites and return samples from the Ryugu asteroid \citep{Hassan2024, Aponte23}, but also recently in the infrared spectrum of comet C/2017 K2 \citep{Woodward2025}. Finally, as complex organic molecules, PAHs are considered potential contributors to prebiotic chemistry, offering essential building blocks for macromolecules relevant to life \citep{Groen2012, Sandford2020}.

One major challenge in studying PAH in space is the difficulty in observing specific PAH molecules directly: AIBs arise from emission of C-C and C-H vibrational modes, which are common to the entire PAH population. Until recently, the detection of C$_{60}$ buckminsterfullerene and its cation \citep{Cami2010, Garcia2010C60,Campbell2015} has been considered the best indirect proof of the presence of large carbonaceous molecules such as PAHs in the ISM. This situation has now changed, thanks to radio observations that identified several PAHs with up to 7 rings in cold molecular clouds \citep{McGuire2021, Burkhardt2021, Cernicharo2021, Wenzel2024,Wenzel2025}.

Despite the relevance of PAHs in the physics and chemistry of the ISM, their origin is not well understood. A plausible formation pathway is through chemical processes in the outflows of evolved carbon-rich stars \citep{Cherchneff}. In fact, the temperature and densities in these environments closely resemble those of combustion environments on Earth, where PAHs are abundantly formed as a by-product or intermediary of soot formation. From pyrolysis experiments, numerous chemical formation pathways to PAHs have been found based on the Hydrogen Abstraction Carbon (C$_{2}$H$_{2}$) Addition (HACA) mechanism \citep[e.g.][]{Miller1992, Frenklach1988, Cole1984, Mukherjee1994, Krestinin2000}. However, existing models that incorporate such formation pathways fail to reproduce observed PAH abundances \citep{Cherchneff1992, Cau2002, Cherchneff}. To complicate the situation further, experimental work on carbon dust analogue formation in conditions similar to those of evolved stars has shown that aromatic structures, such as PAHs, are not the primary product \citep{Martinez2020}, requiring additional mechanisms to explain their origin. Increased effort in this topic is leading to the emergence of new high-temperature gas-phase formation pathways for some medium size PAHs \citep[e.g.][]{Goettl2023}. The recent identification of pyrene and naphthalene molecules in return samples from asteroid Ryugu \citep{Aponte23, Zeichner2023} and of their nitrile-substituted version in cold molecular clouds \citep{Wenzel2024} suggest that some PAHs are formed through gas-phase reactions at very low (<~50 K) temperature. The exact nature of these reactions remains unknown.

Corannulene (C$_{20}$H$_{10}$) is a bowl-shaped PAH that can be seen as a cap of the buckminsterfullere structure. Pyrolysis studies have shown abundant formation of this curved PAH molecule \citep{Lafleur1993} likely due to HACA-type mechanism for its formation \citep{Pope1993, Martin2022}. Recently \citet{Zhao2021} proposed a refined high-temperature gas-phase mechanism for the formation of corannulene starting from phenyl. These studies support the idea that corannulene could be formed in the outflow of carbon-rich stars. Due to its high dipole moment of 2.07~D, the rotational spectrum  of C$_{20}$H$_{10}$ has been experimentally characterized in detail \citep{Lovas2005,Perez2017} and the molecule searched for \citep{Pilleri, moran2025}. Finally corannulene has been detected in the Allende meteorite \citep{Becker1997}.

The Red Rectangle nebula  is the brightest emitter of PAH-AIBs in the Galaxy \citep{Waters1998,Song2007, Candian2012}. It is a pre-planetary nebula, created by the binary system HD~44179. The binary's post-AGB star \citep{Winckel2004} undergoes mass-loss and its outflow is funnelled into a biconical morphology by a circumbinary disk \citep{Waters1998}, creating the nebula's distinct X-shaped, biconical morphology \citep{Cohen}. The nebula has been extensively studied in context of this peculiar morphology, giving insights into the temperature and velocity structures within the system. The nebula has also been widely studied because of its abundance of spectroscopic features of unknown origin, including the Extended Red Emission (ERE) \citep{Schmidt1991, VanWinckel2002}, blue luminescence (BL) \citep{Vijh2004}, and emission bands aligned with the Diffuse Interstellar Bands (DIBs) \citep{Schmidt1991, Sarre1995}. These features were originally thought to be the results of unusual "dual" chemistry \citep{Waters1998,Cohen1999, Guzman2011}, created by the presence of an oxygen-rich disk and a carbon-rich outflow, where PAHs form. The recent detection of both oxygen and carbon molecules (e.g. SiO, H$_2$O and H$^{13}$CN) in the circumbinary disk suggests the presence of a photodissociation region \citep{Bujarrabal2013, Bujarrabal2016, Bujarrabal2023, GallardoCava2022}. Recent ALMA observations furthermore indicate that the X-shaped wind emanates from the disk itself \citep{Bujarrabal2023}, which opens the possibility that PAHs are already present in the disk itself, before being released into nebula's outflow.
Despite the abundance of spectroscopic features, observations show a relative lack of molecules and molecular emission, which is a common feature in binary post-AGB nebulae, and particularly in those in which their rotating circumbinary disks represent most of the nebular mass \citep{GallardoCava2022}.

Previous observations of the Red Rectangle nebula by \cite{Pilleri} using the IRAM-30m telescope provided an upper limit to the fraction of carbon locked in corannulene ($\sim 10^{-5}$). In this paper, we present new high-resolution and high-sensitivity observations of the Red Rectangle nebula, using the Atacama Large Millimeter/submillimeter Array (ALMA) (see section \ref{observations}). Despite the telescope's high sensitivity, and despite boosting the signal through line stacking (see section \ref{methods_linestacking}), we did not detect corannulene but establish a new upper limit of $5 \times 10^{-13}$ compared to hydrogen. 
Additionally, we report tentative detections of emission signals at 139.612 GHz, 139.617 GHz, and 139.621 GHz (see section \ref{results_unknown_emission}). These results are discussed and summarised respectively in sections \ref{discussion} and \ref{conclusions}.

\section{Observations and Data Reduction}
\label{observations}

The observations were taken with ALMA band 4, and were obtained across 5 separate nights, between 26 October 2018 and 3 May 2019 (Project 18.1.01577.S., PI Candian). In total, 11 separate observations were taken, amounting to 8.959 hours of observation time. The 12m array was used with baselines up to $\sim$500 m, resulting in a spatial resolution of 0.419$\arcsec$. The spectral windows spanned the frequencies 136.54-136.65 GHz, 137.55-137.67 GHz, 139.59-139.71 GHz, 149.78-149.90 GHz, 150.80-150.91 GHz with a spectral resolution of 244 kHz. Continuum data was also taken at 150.51-152.49 GHz with a spectral resolution of 31250 kHz.

Six corannulene rotational transitions were targeted:
\begin{itemize}
    \item 137.615 GHz (J+1 → J = 135-134)
    \item 138.633 GHz  (J+1 → J = 136-135) 
    \item 139.652 GHz (J+1 → J = 137-136) 
    \item  149.838 GHz (J+1 → J = 147-146) 
    \item  150.857 GHz (J+1 → J = 148-147)  
\end{itemize} 
 These frequencies where determined by two sets of experiments. \citet{Lovas2005} recorded experimentally the high-resolution rotational spectrum of corannulene up to J=19: the spectrum showed a lack of K-splitting, consistent with corannulene being a rigid molecule. A follow up experiment recorded the frequency of the  112 → 111 transition \citep{Pilleri}. This confirmed the predicted frequency within 250~kHz and also the lack of K-splitting at this high frequency. This means that for corannulene the different K-stacks in which rotational transitions arrange fall on top of each other, increasing the intensities of rotational transitions. More details about the experiments and calculations can be found in the aforementioned papers.

\begin{table}
\centering
\begin{tabular}{llll}
\textbf{} & \textbf{Day} & \textbf{Bandpass Calibrator} & \textbf{Gain Calibrator} \\
1 & 26 Oct 2018 & J0522-3627 & J0607-0834\\
2 & 12 Nov 2018 &J0522-3627 & J0607-0834\\
3 & 12 Nov 2018&  J0522-3627& J0607-0834\\
4 & 12 Nov 2018 & J0522-3627& J0607-0834\\
5 & 14 Dec 2018 &J0522-3627 & J0607-0834\\
6 & 14 Dec 2018    &    J0522-3627  & J0607-0834\\
7 &  14 Dec 2018    &  J0522-3627    & J0607-0834\\
8 & 2 May 2019  & J0725-0054 & J0607-0834\\
9 & 2 May 2019  & J0725-0054 & J0607-0834 \\
10 & 2 May 2019  & J0725-0054     &J0607-0834 \\
11 & 3 May 2019 & J0725-0054 & J0607-0834\\
\end{tabular}
\caption{: Overview of the ALMA observations and the used calibrator objects.}
\label{tab:calibrator_overview}
\end{table}
\newcommand{\seca}{\mbox{\rlap{.}$''$}}

Standard calibration was performed with the CASA 5.4.0-70 pipeline, using the quasars outlined in Table \ref{tab:calibrator_overview} as bandpass and gain calibrators. Subsequently, self-calibration was performed on all datasets, using continuum data self-calibration solutions. The resulting continuum image has a synthetic beam with Half Power Beam Width (HPBW) of $\sim 0.5\arcsec\times0.6\arcsec$ and shows an oval emission area of $\sim 1.3\arcsec\times2.5\arcsec$ with flux $\sim80$ Jy, where we probe the central parts of the Red Rectangle.

Image synthesis of the line data was done with the CASA version 5.80, using the Hogbom method with a natural weighting scheme. The resulting image cubes have a spectral resolution of 488 kHz and a synthetic beam with Half Power Beam Width (HPBW) of $\sim$ 0\seca7$\times$0\seca8.

\section{Methods}
\subsection{Spectral Line Stacking}
\label{methods_linestacking}
Spectral stacking has previously proven effective in amplifying weak signals \citep[e.g.,][]{McGuire2021}. We applied this method to our six spectral windows, each targeting one of the six observed corannulene transitions. To perform the stacking, we used the Python package \textit{LineStacker} \citep{Jolly2020}, which integrates Python and CASA tasks to stack extracted spectra or full 3D image cubes.

Using a similar procedure as \citet{McGuire2021}, we stacked the ALMA data cubes according to the local noise level. We use a $\sigma$-weighted averaging approach, where each pixel is averaged across the six datasets. This process was carried out independently for each spectral channel, with velocity channels binned at 1 km s$^{-1}$ and the various spectra aligned according to these velocity bins. Hereby we expect to be minimally impacted by the frequency uncertainties of 250~kHz, which corresponds to a velocity uncertainty of $\sim$0.2-0.5~km s$^{-1}$, depending on the spectral window. \citet{McGuire2021} furthermore apply weighing according to the expected brightness level of each line. However, since the targeted spectral lines are expected to exhibit similar brightness in the Red Rectangle nebula \citep[see Fig. 2 in ][]{Pilleri}, we did not apply weighting.

\subsection{Upper Limit Derivation}
\label{methods_upperlimit}
To determine an upper limit on the corannulene abundance, we followed the derivation outlined in \citet{Pilleri}, which involves calculating the ratio $\frac{F_{3\sigma}}{F_{\text{model}}}$. Here, $F_{3\sigma}$ represents the maximum flux of corannulene emission that remains undetected by our observations. $F_{\text{model}}$ is the expected flux from all PAHs, as modelled by \citet{Mulas2006_feb}. 

The expected corannulene emission ($F_{\text{model}}$) is $19\times10^{-21} $ W cm$^{-2}$ for all relevant emission frequencies \citep[See Fig. 2 in ][]{Pilleri}. To calculate $F_{3\sigma}$, we obtain spectra of an assumed corannulene emission region. We inferred this emission region to coincide with the Aromatic Infrared Bands (AIBs) emission region. The AIBs were observed by the Infrared Space Observatory (ISO) using the Short Wavelength Spectrometer (SWS), revealing a rectangular emission area of $10\arcsec\times20\arcsec$ \citep{Waters1998}. Using the imaging tool CARTA, we obtained the spectra for such a rectangular area, centred on the inner binary and oriented perpendicularly to the equatorial disk \citep{Bujarrabal2023}. Note that the Maximum Recoverable Scale of our ALMA data is 10", and thus our sensitivity is limited to emitting regions smaller than this scale. This aligns well with the more compact size of the disk emission \citep{Bujarrabal2016} and localized PAH emission regions shown in \citet{Waters1998} to be distributed within the $10\arcsec\times20\arcsec$ area.

Finally, we stack the extracted emission spectra by averaging them together without weighting. The 3$\sigma$ level of the stacked spectrum then simply indicates the upper limit for observing all six stacked lines, as the lines contribute (within our order of precision) equally to the maximum flux \citep[][Fig 2]{Pilleri}. To obtain the maximum emission flux of all lines ($F_{3\sigma}$), we assume that the lines are broadened with a FWHM of 1 km~s$^{-1}$.

\section{Results}
\subsection{Upper Limit on the Abundance of Corannulene and Small PAHs in the Red Rectangle Nebula}
\label{result_upperlim}
None of the ALMA images (see appendix \ref{appendix_nondetections}) showed emission that aligns with the rest frequency of corannulene, taking into account the systemic velocity of the system \citep[V$_{LSR}$ = -0.25 km s$^{-1}$,][]{Bujarrabal2016}. From the observed  $^{12}$CO J = 6-5 transition velocities up to $\pm$ $\sim$10 km s$^{-1}$ are estimated \citep{Bujarrabal2016}, we expect similar values for corannulene. Considering that in the Red Rectangle velocities can be up to $\sim$200 km/s in a jet inclined along the outflow \citep{Thomas2013}, we can even argue that radial velocities can range up to $\sim$25 km/s \citep[taking i = 85\textdegree, ][]{Thomas2013}. However, no clear emission signal was found in any of these velocity ranges either. Spectral line stacking (section \ref{methods_linestacking}) likewise did not result in a detection of corannulene (see appendix \ref{appendix_nondetections}).

An upper limit was derived, using the ratio $\frac{F_{3\sigma}}{F_{\text{model}}}$ (section \ref{methods_upperlimit}). We measured $F_{3\sigma} = 4.2 \times 10^{-27}$ W cm$^{-2}$ for the lines together. With $F_{\text{model, line}} = 19 \times 10^{-21}$ W cm$^{-2}$ for each separate line, which gives a total flux of $ F_{\text{model, total}} =  6\times F_{\text{model, line}}$ for the six lines together, we find $\frac{F_{3\sigma}}{F_{\text{model, total}}} = 3.7 \times 10^{-8}$. However, we should account for corannulene molecules with a rotational spectrum different to the spectrum targeted by our observations. First, a $^{13}$CO/$^{12}$CO ratio of $\sim 0.1$ has been estimated in the Red Rectangle \citep{Bujarrabal2016}, implying that for every one of the 20 carbon molecules in corannulene there is a 1/11 probability of $^{13}$C substitution. From this we calculate the probability of having at least one  $^{13}$C substition, versus the probability of containing only $^{12}$C. We find a $^{13}$C-substituted/$^{12}$C-only ratio of $\sim$5.7/1. Additionally, 60\% of corannulene molecules in this environment are expected to be in the vibrational ground state, with 40\% in the vibrationally excited states \citep{Pilleri}. These molecules have a distinct rotational spectrum from the pure, ground-state corannulene molecules, and thus would not be detected by these ALMA observations. 

Taking these considerations into account, we correct the flux ratio to account for all corannulene molecules. Finally, we derive  an upper limit of $4 \times 10^{-7}$ for the fraction of carbon in corannulene relative to the total amount of carbon contained in all PAHs within the Red Rectangle. This is $1-2$ orders of magnitude lower than previous estimates by \citet{Pilleri}, which can be attributed to the higher sensitivity of ALMA and the use of spectral line stacking. Converting this to the fraction of corannulene compared to elemental carbon or hydrogen, and assuming 10\% of cosmic carbon is locked in PAHs \citep{Draine2007} with a C/H ratio of $3.39 \times 10^{-4}$ \citep{Zuo2021}, we estimated the upper limit for the fraction of \textbf{elemental carbon locked in corannulene to be $4 \times 10^{-8}$}. This translates to an upper limit for the corannulene-to-hydrogen ratio: $\frac{N_{C_{20}H_{10}}}{N_H} = 5 \times 10^{-13}$.

\begin{figure}
    \centering 
    \includegraphics[width=0.49\textwidth]{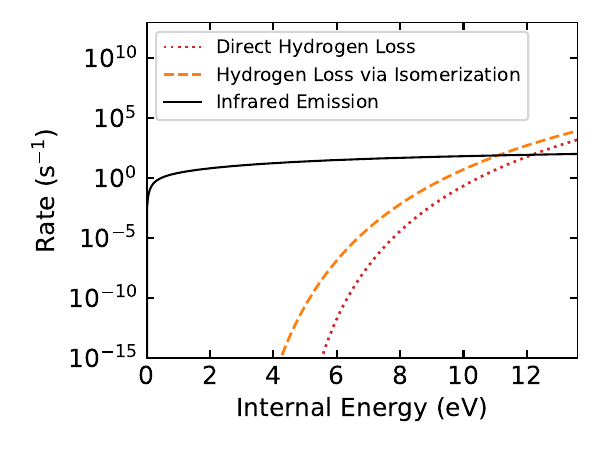}
    \caption{The photodissociation rate for the loss of hydrogen directly from corannulene cation (in red) or via an isomerized state of corannulene (in orange). The infrared emission rate of the corannulene neutral is shown in black.}
    \label{fig:finalrates}
\end{figure}

 \begin{figure}
    \centering
    \includegraphics[width=0.49\textwidth]{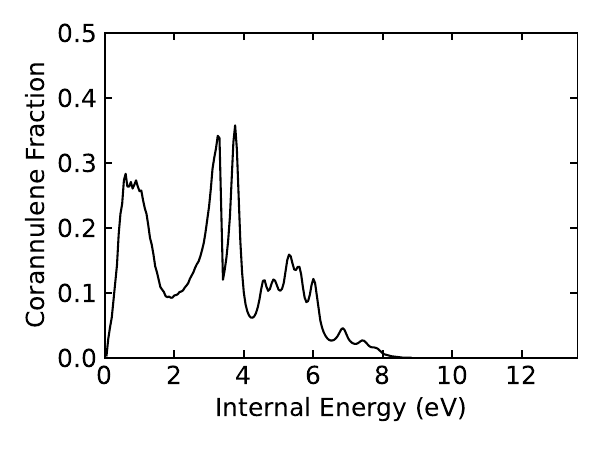}
    \caption{Calculation of the internal energy distribution of the corannulene molecules in the Red Rectangle radiation field.}
    \label{fig:E_distribution}
\end{figure}

\subsection{Photodissociation of corannulene in the Red Rectangle}
\label{result_photodissociation}
Motivated by the lack of small PAHs in the Red Rectangle, we investigate the potential for photodissociation of corannulene in such an environment. From experiments and computational chemistry calculation we know that  the lowest dissociation channel for corannulene cations is H-loss \citep{West2018, Sundararajan2024}. In Appendix~\ref{methods_photodissociation} we summarise the finding of these work and describe how we calculate photodissociation rates for corannulene, which are presented in Figure~\ref{fig:finalrates}.  

  The hydrogen loss via an isomerized state is the dominant relaxation mechanism for corannulene up to internal energies of at least 13.6 eV. The infrared emission rate of the neutral corannulene is also depicted, and follows a flatter curve reaching rates of $\sim10^2$ s$^{-1}$. Up to $\sim$11 eV (10~eV if considering the modelling by \cite[]{West2018}) the infrared emission is dominant over both of the photodissociation channels.

Since these reaction rates are dependent on the internal energy, we consider now the internal energy distribution of corannulene in the Red Rectangle environment, shown in figure \ref{fig:E_distribution} (see derivation in appendix \ref{methods_photodissociation}). The corannulene molecules do not reach internal energies larger than $\sim9$ eV. By comparing this energy distribution with the photodissociation rates (figure \ref{fig:finalrates}), it is clear that the photodissociation of corannulene is unlikely to happen in the Red Rectangle: the internal energy distribution reaches zero for energies above $\sim9$ eV, while it is only above $\sim10-11$ eV that photodissociation dominates over the infrared emission.  Absorption of multiple photons is needed to dissociate large PAH molecules \citep{Montillaud2013}. However, for molecules of the size of corannulene, \citet{Andrews2016} (their Fig. 6) shows the importance of multi-photon absorption is marginal, even at G$_0$ values of some 10$^5$ calculated for the Red Rectangle \citep{Fong2001}.

\subsection{Unidentified emission around 139.617 GHz and 139.632 GHz}
\label{results_unknown_emission}
\begin{figure*}
    \centering
    \includegraphics[width=0.6\textwidth]{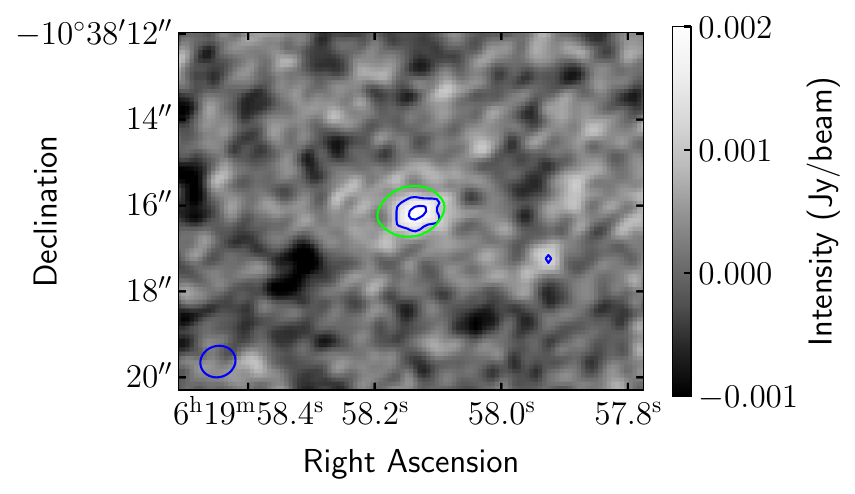}
    \caption{The Red Rectangle as observed at a frequency of 139.617~GHz. The blue contour shows the $3\sigma$ and $5\sigma$ line emission region, while the green contour shows the $3\sigma$ continuum emission region. The blue ellipse in the bottom-left corner shows the image beam.}
    \label{fig:unknown_emission_image}
\end{figure*}

\begin{figure*}
    \centering
    \includegraphics[width=0.7\textwidth]{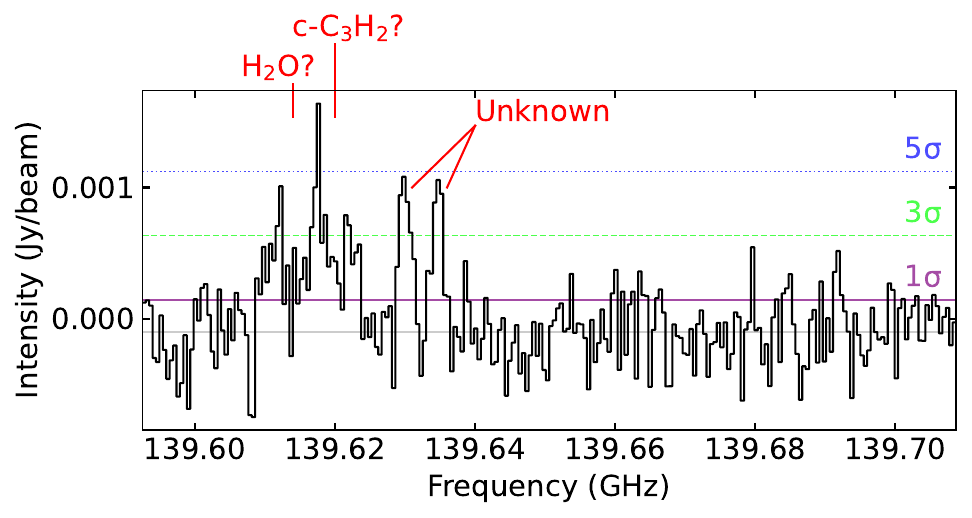}
    \caption{The spectrum corresponding to the 139.617~GHz emission region, shown in figure \ref{fig:unknown_emission_image}. The $1\sigma$, $3\sigma$, and $5\sigma$ levels are shown, with the brightest emission peak reaching above $5\sigma$, while four other peaks span ranges of $\sim 3\sigma - 5\sigma$. The origin of two of these emission peaks remains unknown. The other the peaks could be a result of the 140~GHz water maser and cyclopropenyledine (c-C$_3$H$_2$), whose rest frequencies are indicated.}
    \label{fig:unknown_emission_spectrum}
\end{figure*}

We observe an emission signal at 139.617 GHz, shown in figure \ref{fig:unknown_emission_image}. The spectrum corresponding to this emission region (see figure \ref{fig:unknown_emission_spectrum}) furthermore shows several  $\sim 3-5 \sigma$ peaks at frequencies of 139.612, 139.621, 139.630, and 139.634 GHz. However, there is no clear emission region in these frequencies, as shown in figure \ref{fig:other_emission_images}. The gain calibrator spectra were inspected to rule out an improper bandpass calibration, and the rms noise is stable across these frequencies, suggesting that these peaks cannot be due to atmospheric noise. Thus, we report a tentative detection.

The brightest peak at 139.617 GHz would correspond to a corannulene velocity of $\sim75$km s$^{-1}$. This is not a velocity expected to be seen in the Red Rectangle nebula (see section \ref{result_upperlim}) and a signal of similar velocity does not appear for any of the other spectral windows. Since the observed emission lines are expected to be equally bright \citep[see figure 2 in ][]{Pilleri}, this emission therefore does not seem to be a result of corannulene. We further discuss the possible origin of these emission feature in section \ref{discussion_unknown_emission}.

\section{Discussion}
\label{discussion}
\subsection{The lack of corannulene in the Red Rectangle nebula}
\label{discussion_upperlim}
The ALMA observations show that corannulene is not detected in the Red Rectangle and we derived a firm upper limit for its abundance of $5\times10^{-13}$ with respect to hydrogen. In contrast, C$_{60}$ and C$_{70}$, the only large carbonaceous molecules identified in evolved stars, have abundances of the order of 10$^{-8}$ with respect to hydrogen (see Table \ref{tab:abundances}). In the following we discuss different possiblities for this scenario.

\begin{description}
    \item[Destruction:]  Our calculations (see Sect.~\ref{result_photodissociation}) for H-loss based on the experimentally observed photodissociation behaviour of the corannulene cation confirmed earlier estimates by \citet{Pilleri} that the molecule should be stable under the radiation field of the Red Rectangle. \citet{Sundararajan2024} showed that carbon-loss  can also happen if enough energy is deposited in corannulene. This energy could in principle come from energetic protons in the stellar wind \citep{Chanpeaux2014}.
    Expanding the observational coverage of the Red Rectangle’s radiation field to higher energy ranges (9–13.6 eV) \citep{Vijh2005} could help confirm the absence of high-energy photons that might enhance photodissociation rates.
    \item[Formation:] Our understanding of PAH formation in evolved stars at high temperature come from analogy with Earth pyrolysis study. Corannulene is created in high abundance through HACA-type mechanism \citep{Lafleur1993, Pope1993}. Recently a new mechanism to create the molecule at high temperature has been proposed  \citep{Zhao2021}. It would be interesting to test it in an astrochemical model for carbon-rich evolved stars. Corannulene might also be formed through a low-temperature gas-phase route, as is proposed for coronene after its detection in TMC-1 \citep{Wenzel2025}. However, several attempts to search for corannulene in the same molecular cloud have been so far unsuccessful \citep{moran2025}, and so the low-temperature route seems unlikely. Corannulene has been detected in the Allende meteorite \citep{Becker1997} but no information about its carbon isotopic compositions are available. Further detection of corannulene in meteorites or cold ISM environments would be beneficial in assessing if and how this PAH can be formed.
    \item[Isotopologues:] The measured $^{13}$C/$^{12}$C ratio in the Red Rectangle is 0.1 (Sect.~\ref{result_upperlim}). If we assume that corannulene is efficiently created in the Red Rectangle and this isotopic ration, then $\sim$85\% percent of corannulene molecules would contain at least one $^{13}$C. The  substitution of one C reduces the symmetry of the molecule from an oblate symmetric top to that of a near-oblate asymmetric top. \citet{Perez2017} recorded the broad band rotational spectrum of $^{13}$C substituted corannulene between 3000 and 7500 MHz.  The lower symmetry of the substituted corannulene leads to different amount of K-type splitting in the spectra, depending on the position where the atom is substituted. Further experiments at high frequency would be helpful in constraining the exact position of the rotational transitions for $^{13}$C substituted corannulene and allow for a search of this molecule.
\end{description}

\subsection{Tentative Detection of 140 GHz water maser and c-C$_3$H$_2$}
\label{discussion_unknown_emission}
In Section \ref{results_unknown_emission}, we reported the detection of an unexpected emission feature within the 139.612 - 139.634~GHz range (Figs. \ref{fig:unknown_emission_image} \& \ref{fig:unknown_emission_spectrum}). This emission is unlikely to originate from corannulene, given the improbably high associated velocities ($\sim$75 km s$^{-1}$) and the absence of similar features in other spectral windows. To investigate alternative molecular sources, we utilized the Splatalogue Database for Astronomical Spectroscopy\footnote{\href{https://splatalogue.online}{https://splatalogue.online}}.

The features observed at 139.612 GHz, 139.617 GHz, and 139.621 GHz could be attributed to cyclopropenylidene (c-C$_3$H$_2$) and water within the circumbinary disk (see Table \ref{tab:detected_molecules}). In this scenario, the radial motion of the disk would result in the splitting of these rest frequencies into distinct blue- and red-shifted components. These components subsequently overlap to create a pronounced emission peak at 139.617 GHz, flanked by weaker peaks at 139.614 GHz and 139.620 GHz. The inferred velocity shifts of ±6–7 km s$^{-1}$ are consistent with previously observed rotation velocities of the circumbinary disk, which in most cases also reach velocities of ±6–7 km s$^{-1}$ \citep{Bujarrabal2023}. No other molecules in the Splatalogue database are identified as possible candidates for the observed emission. It could also be that molecules for which rotational spectra are unknown are responsible for these features. Although further observations are needed to confirm, we thus report a tentative detection of these molecules.

The first molecule, cyclopropenylidene (c-C$_3$H$_2$), is a cyclic, partially aromatic hydrocarbon. Although it spatially overlaps with the oxygen-rich disk of the Red Rectangle environment, recent work by \citet{VandeSande2024} demonstrates that such carbon-rich molecules can form even in oxygen-rich disk environments. This molecule has also been identified previously in PDRs. The co-spatial distribution of c-C$_3$H$_2$ with the disk, along with a possible double-peaked emission pattern, suggests a connection to the PDR region proposed by \citet{Bujarrabal2013}. In PDRs, the molecule's formation has been linked to PAH evolution \citep{Pety2005, Fuente2003}, although reaction pathway models fail to reproduce its formation \citep{Murga2020}. Thus, the detection of c-C$_3$H$_2$ would serve as a new tracer for the Red Rectangle PDR, and offer valuable insights into the environment's hydrocarbon chemistry. Whether c-C$_3$H$_2$ originates from PAH fragmentation or through a bottom-up chemical process remains unclear, highlighting the need for detailed modelling to explore these pathways.

To investigate the potential presence of c-C$_3$H$_2$ in the Red Rectangle, we cross-referenced all available ALMA data with the spectral lines of c-C$_3$H$_2$ listed in the Splatalogue database, and falling  within the relevant frequency ranges (as listed on the ALMA data archive). While most of the ALMA datasets lacked the sensitivity required, several lines within the frequency range of ALMA Cycle~0 project 2011.0.00223.S were identified as potentially detectable, where the SNR was estimated to be $\sim1.5$x higher than the c-C$_3$H$_2$ line in our dataset. To further investigate this tentative detection, future work could involve collecting the available ALMA data on the Red Rectangle and applying line stacking techniques and/or LTE modeling to confirm or disprove the presence of c-C$_3$H$_2$. However, such an analysis is beyond the scope of the current work.

\begin{table}   
\centering
\caption{Properties of the tentatively detected lines c-C$_3$H$_2$ and H$_2$O (from the Splatalogue Database).}
\begin{tabular}{lll}
\textbf{} & \textbf{c-C$_3$H$_2$} & \textbf{H$_2$O} \\ \hline
\textbf{Rest Frequency (GHz)} & 139.6201116 & 139.614293 \\
\textbf{Resolved Quantum Numbers} & 13(11, 3)-13(10, 4) & 14( 6, 9)-15( 3,12) \\
\textbf{Upper Level Energy (K)} & 286.28565 & 4438.39649
\end{tabular}

\label{tab:detected_molecules}
\end{table}

Water maser emission has already been observed in the circumbinary disk of the Red Rectangle, namely the H$_2$O $J_{K_{\mathrm{a}}, K_{\mathrm{c}}}=3_{2,1}-4_{1,4}$ transition at 331.12 GHz \citep{Bujarrabal2023}. Therefore, it is plausible that the 140 GHz water maser line has been detected in our observations. If this is confirmed by further observations, this might represent the first detection of the 140 GHz water maser line \citep{Gray2015}, offering new insights for radiative transfer models of water masers. Notably, \citet{Gray2015} find this transition is only present as a maser under conditions of high dust temperature ($T_\mathrm{d} = 1025$ K), but not at lower temperatures (e.g., $T_\mathrm{d} = 50$ K). High temperatures of $T_\mathrm{d} \approx 1000$ K have been inferred for the innermost part of the disk \citep{Menshchikov2002}, furthermore supporting the possibility of such maser emission.

We note that another H$_2$O $\nu$ = 0 maser was previously detected in the Red Rectangle (at 22 GHz) \citep{GallardoCava2022}, showing redshifted profile that peaks at 8.3 km/s, instead of the double-peaked profile typical for a disk. This asymmetry may result from selective excitation affecting only one side of the disk. A similar phenomenon could also occur in the 140 GHz water maser emission observed here. However, our spectrum shows that the central peak at 139.617 GHz is more prominent than the flanking peaks at 139.612 GHz and 139.621 GHz. This suggests that both c-C$_3$H$_2$ and H$_2$O contribute to a two-peak profile, with their combined emission enhancing the central feature. Nevertheless, we cannot rule out the possibility that asymmetric excitation affects the 140 GHz water maser line.

Finally, the unidentified features at 139.630 GHz and 139.634 GHz lie on either side of a methanol ($\mathrm{CH_3OH}$) transition. No other methanol lines have been reported in the Red Rectangle and methanol is not commonly found in circumstellar envelopes in general. It however has been detected in a few oxygen-rich sources exhibiting signs of active shocks \citep{SanchezContreras2018} and so the presence of a transition near these frequencies is worth noting. Also in this scenario, if the emission arises from the inner regions of the rotating disk, the line could be split to create the features seen at 139.630 GHz and 139.634 GHz. Although the upper energy level of this methanol transition is relatively high ($\sim$3000 K), it is a $v=2$ vibrationally excited line and could be excited via infrared pumping, as has been suggested for other detected transitions in this source, such the H$_2$O ($v=2$) line \citep{Bujarrabal2023}. However, a more robust analysis is required before even a tentative detection can be claimed.

\section{Summary and Conclusion}
\label{conclusions}
We presented high-resolution ALMA observations of the Red Rectangle nebula, aimed at detecting the specific PAH molecule corannulene (C$_{20}$H$_{10}$). Although we did not detect corannulene emission, we were able to derive an upper limit on its abundance. We discussed the lack of detection in the context of PAH formation and destruction mechanism.

Furthermore, we report tentative detections of emission features potentially originating from cyclopropenylidene (c-C$_3$H$_2$) and the 140 GHz water maser line. The identification of c-C$_3$H$_2$ would provide new insights into the hydrocarbon chemistry of the Red Rectangle, as well as serve as a tracer for the system's photodissociation region (PDR). Similarly, the detection of the 140 GHz water maser could advance our understanding of water maser emission. To better understand the Red Rectangle's chemical environment in relation to PAH formation, as well as its supposed PDR-like chemistry, follow-up observations of these molecules in other bands are recommended.

In conclusion, this study offers valuable insights into the intricate chemical environment of the Red Rectangle nebula, revealing the potential presence of c-C$_3$H$_2$ and the 140 GHz water maser. These findings challenge existing models of PAH abundance and formation, suggesting that updated theoretical frameworks are necessary to better understand PAH chemistry in post-AGB environments.

\section*{Acknowledgements}
The authors acknowledge Giacomo Mulas for fruitful discussion and Allegro, the European ALMA Regional Center node in the Netherlands for assistance with data analysis.

J.A and V.B are partially supported by the I+D+i projects PID2019-105203GB-C21 and PID2023-146056NB-C21, funded by the AEI (10.13039/501100011033) of the Spanish MICIU and the European Regional Development Fund (ERDF) of the EU.

J.C. acknowledges support from the Natural Sciences and Engineering Research Council (NSERC) of Canada.

ALMA is a partnership of ESO (representing its member states), NSF (USA) and NINS (Japan), together with NRC (Canada), NSTC and ASIAA (Taiwan), and KASI (Republic of Korea), in cooperation with the Republic of Chile. The Joint ALMA Observatory is operated by ESO, AUI/NRAO and NAOJ. 

This article is based upon work from COST Action CA21126 - Carbon molecular nanostructures in space (NanoSpace), supported by COST (European Cooperation in Science and Technology).

%%%%%%%%%%%%%%%%%%%%%%%%%%%%%%%%%%%%%%%%%%%%%%%%%%
\section*{Data Availability}
\label{data}
This paper makes use of the following ALMA data: ADS/JAO.ALMA\#18.1.01577.S. The original ALMA data can be found on the \href{https://almascience.eso.org/aq/?sourceNameResolver=red%20rectangle&observationsPiName=pierre&observationsSortProp=continuumSensitivity&observationsSortDir=asc&result_view=projects&projectsCode=18.1.01577.S}{ALMA archive}, under project code 18.1.01577.S. The CASA scripts used to calibrate these can be found on \href{https://zenodo.org/}{Zenodo}, with DOI 10.5281/zenodo.15404918. Here, we also show the resulting ALMA cubes and all other data and python scripts used to reproduce the figures in this paper.

%%%%%%%%%%%%%%%%%%%% REFERENCES %%%%%%%%%%%%%%%%%%

% The best way to enter references is to use BibTeX:

\bibliographystyle{mnras}
\bibliography{example} % if your bibtex file is called example.bib

% Alternatively you could enter them by hand, like this:
% This method is tedious and prone to error if you have lots of references
%\begin{thebibliography}{99}
%\bibitem[\protect\citeauthoryear{Author}{2012}]{Author2012}
%Author A.~N., 2013, Journal of Improbable Astronomy, 1, 1
%\bibitem[\protect\citeauthoryear{Others}{2013}]{Others2013}
%Others S., 2012, Journal of Interesting Stuff, 17, 198
%\end{thebibliography}

%%%%%%%%%%%%%%%%%%%%%%%%%%%%%%%%%%%%%%%%%%%%%%%%%%

%%%%%%%%%%%%%%%%% APPENDICES %%%%%%%%%%%%%%%%%%%%%

\appendix

% \section{Some extra material}

% If you want to present additional material which would interrupt the flow of the main paper,
% it can be placed in an Appendix which appears after the list of references.

\section{ALMA Image Maps}
\label{appendix_nondetections}
\begin{figure*}
    \centering
    \includegraphics[width=0.95\textwidth]{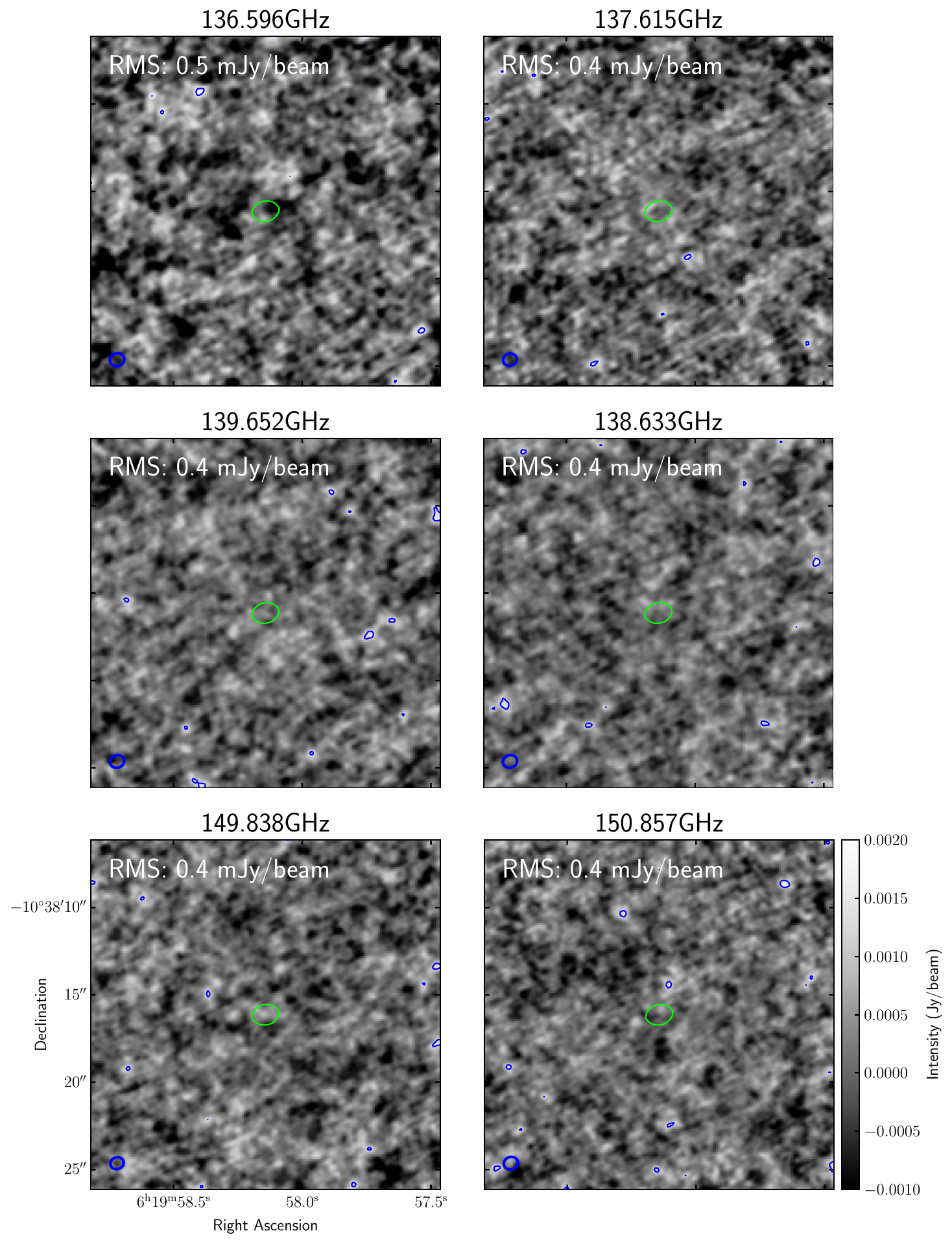}
    \caption{The ALMA images at the rest frequencies of corannulene, where blue contours denote 3$\sigma$ regions and the green contour shows the 3$\sigma$ continuum emission region. The blue ellipse in the bottom-left corner shows the image beam. We also report the corresponding rms noise.}
    \label{fig:nondetections}
\end{figure*}

\begin{figure*}
    \centering
    \includegraphics[width=0.7\textwidth]{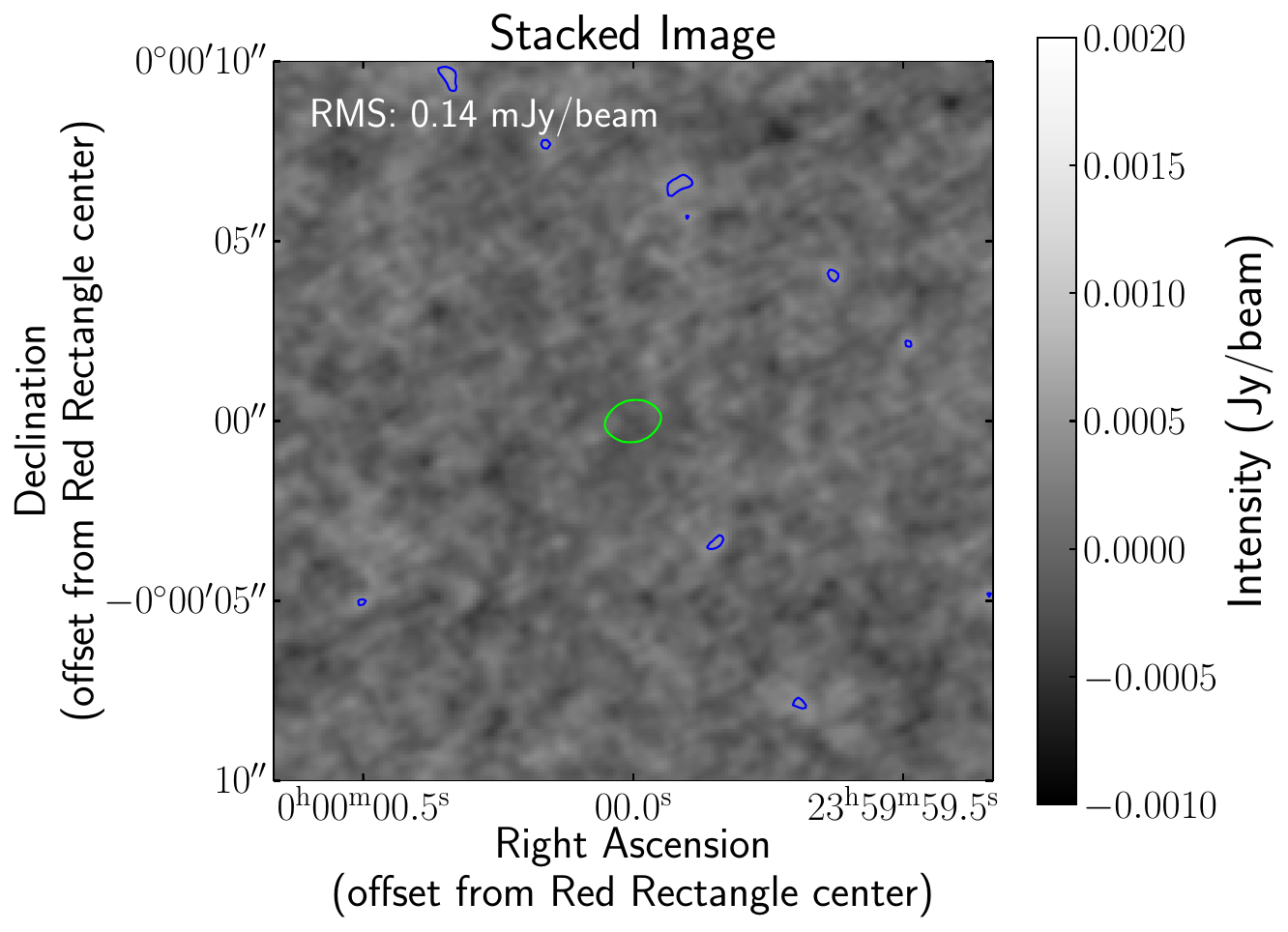}
    \caption{The ALMA images, stacked at the rest frequencies of corannulene, where blue contours denote 3$\sigma$ regions and the green contour shows the 3$\sigma$ continuum emission region. We also report the corresponding rms noise.}
    \label{fig:stacked}
\end{figure*}

\begin{figure*}
    \centering
    \includegraphics[width=0.9\textwidth]{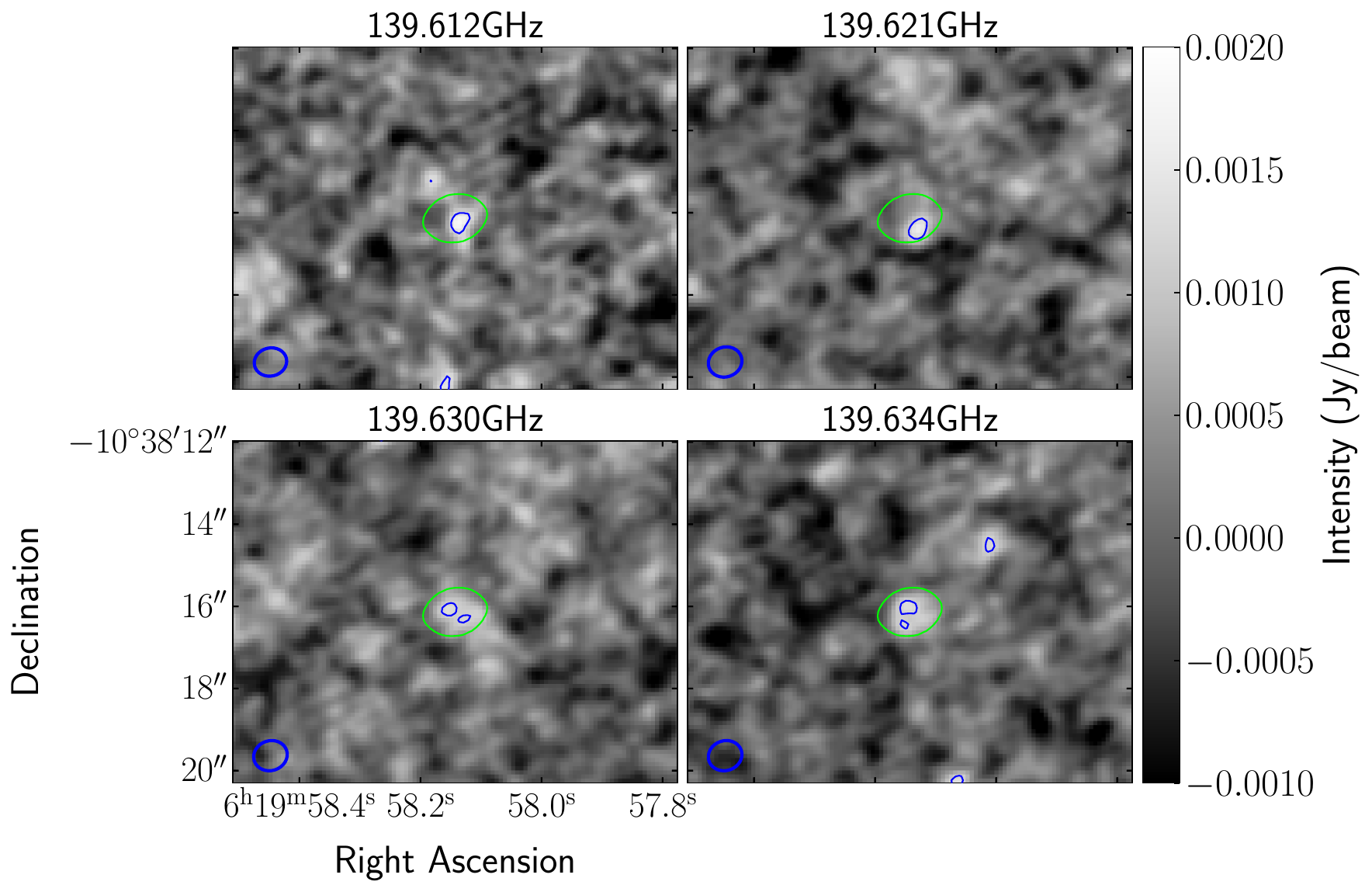}
    \caption{The Red Rectangle as observed at frequencies corresponding to the various 3-5 $\sigma$ emission peaks shown in figure 6. The blue contour shows the $3\sigma$ and $5\sigma$ line emission region, while the green contour shows the $3\sigma$ continuum emission region. The blue ellipse in the bottom-left corner shows the image beam.}
    \label{fig:other_emission_images}
\end{figure*}

The (stacked) ALMA images at the rest frequency of corannulene are shown in Figures \ref{fig:stacked} and \ref{fig:nondetections}, revealing no clear evidence of corannulene emission. This holds true for all frequencies within a $\sim 25$ km s${-1}$ shift from the rest frequency. Figure \ref{fig:other_emission_images} presents the ALMA images at other observed spectral emission peaks (see figure \ref{fig:unknown_emission_spectrum}). Despite the presence of peaks in the spectrum at these frequencies, the corresponding emission regions are not clearly visible in the ALMA images. The full data cubes are available at \href{insert link}{[insert link]}.

\newpage

\section{Calculations of photodissociation rate and infrared emission}
\label{methods_photodissociation}
\begin{figure}
    \centering
    \includegraphics[width=0.4\textwidth]{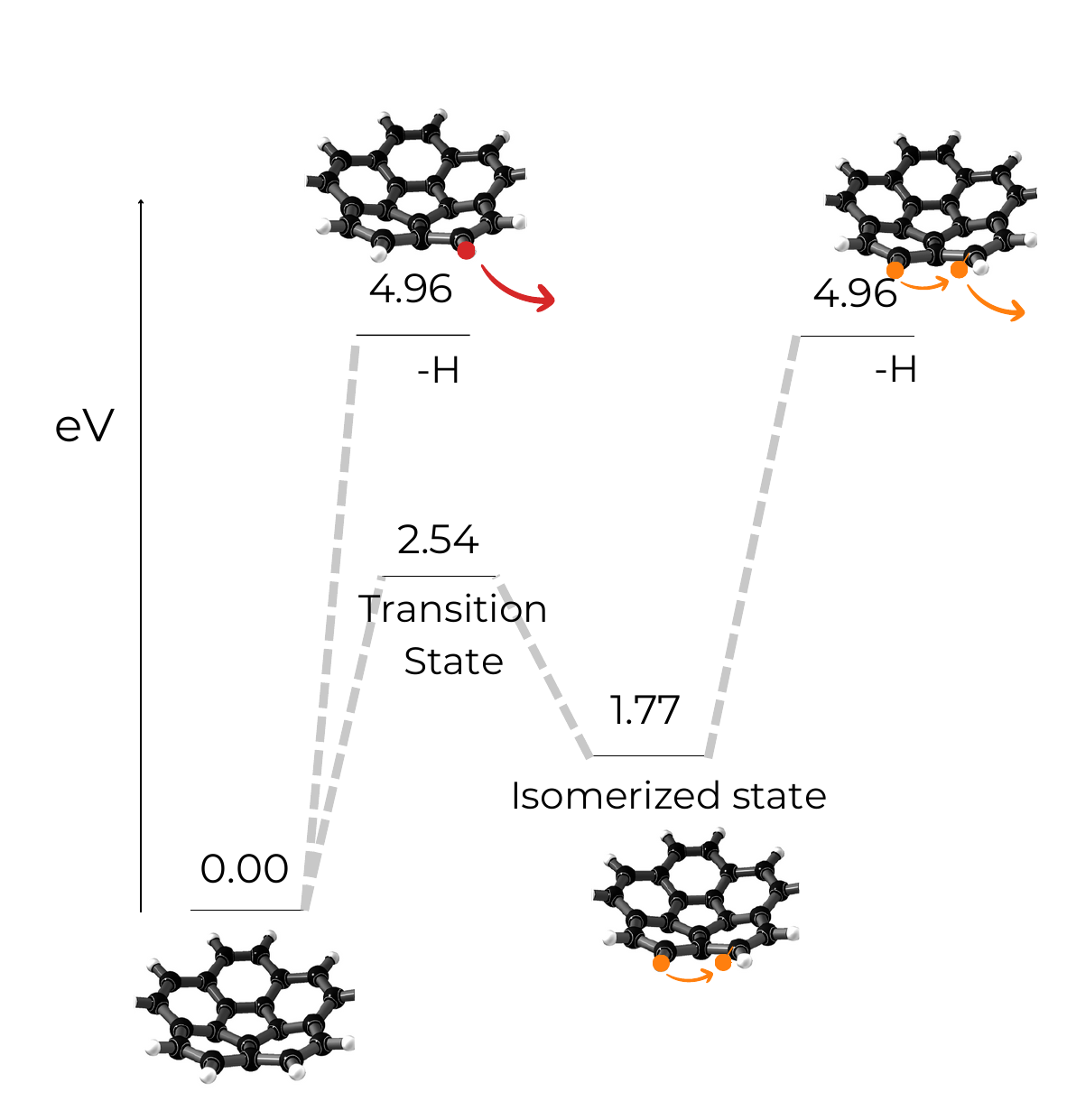}
    \caption{Simplified potential energy surface of the corannulene cation, as calculated by Sundararajan \citep{Sundararajan2024}, used to determine the photodissociation rate of corannulene in the Red Rectangle environment (see Sect. \ref{result_photodissociation}).}
    \label{fig:PES}
\end{figure}

\citet{West2018} determined that the lowest dissociation channel for corannulene cations is H-loss, with
a derived dissociation threshold energy E$_0$=~4.24$\pm$1.03 eV. This behaviour was confirmed recently by \citet{Sundararajan2024} who studied the possible dissociation channels as function of the internal energy using combined experiments and computational chemistry calculations. They found that the H-loss can happen via both direct C-H bond break or through an isomerised intermediate (see Fig.~\ref{fig:PES}). The dissociation requires E$_0 = $ 4.96 eV, consistent with the experimental value obtained by \citet{West2018}.
The total photodissociation rate was calculated by assuming a steady-state approximation for the isomerized state, which is valid, because the isomerized state does not accumulate ($k_{\text{isomerization}} << k_{\text{isomerization}}^*+k_{\text{ -H, from isomerized state}}$). Since many PAH cations have previously shown to behave similarly to their neutral counterparts \citep[e.g.][]{Paris2014}, we assume that these results hold for the corannulene neutral as well. 

We calculate the rate of the uni-molecular reactions involved in corannulene's PES, using RRKM theory \citep{DiGiacomo2014}: 
\begin{equation}
    k_{RRKM}(E) = \frac{W^*(E-E_0)}{h\rho(E)}.
\end{equation}
 Here, $\rho(E)$ refers to the density of state of the molecule with internal energy $E$. $W^*(E-E_0)$ refers to the sum of states for the molecule's transition state, where $E_0$ is the threshold energy for the reaction to occur. The density and sum of states were obtained from the vibrational modes of the various corannulene states \citep{Sundararajan2024}, with the Multiwell tool Densum \footnote{\href{https://multiwell.engin.umich.edu/}{https://multiwell.engin.umich.edu/}}.

The infrared emission was calculated by considering the PAH molecule to be in thermal equilibrium with its surroundings at a temperature $T$ \citep{Pech2002}, and the molecule's vibrational modes to be harmonic oscillators. In this case:
\begin{equation}
\label{k_ir_formula}
    k_{\mathrm{IR}}^i=A_i^{1,0} \times\left[\exp \left(h \nu_i / k T\right)-1\right]^{-1},
\end{equation}
where $\nu_i$ is the vibrational frequency of the bond, $A_i^{1,0}$ is the corresponding Einstein coefficient, and $T$ is the temperature of the molecule. The temperature and internal energy of the molecule ($E$) are related via \citep[][]{Pech2002}: 
\begin{equation}
    E(T)=\sum_{i=1}^s \frac{h \nu_i}{\exp \left(h \nu_i / k T\right)-1}.
\end{equation}

The Einstein coefficients was calculated with the formula \citep{Cook1998}:
\begin{equation}
    A\left(\mathrm{~s}^{-1}\right) \cong\left(1.2512 \times 10^{-7} \frac{\mathrm{mol}~  \mathrm{cm}^2}{\mathrm{~km} \mathrm{~s}}\right) \tilde{v}^2 S\left(\frac{\mathrm{km}}{\mathrm{mol}}\right),
\end{equation}
where $\tilde{v}$ is the wavenumber of the vibrational modes, and $S$ is the line emission intensity. Both the wavenumbers ($\tilde{v}$) and line emission intensities ($S$) of the neutral corannulene molecule were taken from the PAH spectral database\footnote{\href{http://astrochemistry.ca.astro.it/database/corannulene/corannulene\_vib.html}{http://astrochemistry.ca.astro.it/database/corannulene/corannulene\_vib.html}} \citep{Malloci2007}.

Both the photodissociation and infrared emission of the molecule depends on its internal energy. We assume that the internal energy distribution, $n(E)$, is directly determined by the molecule's photon absorption distribution:
\begin{equation}
    R_{\mathrm{abs}}(E)= \sigma_E \frac{F_E}{E},
\end{equation}
where $\sigma_E$ is the absorption cross-section of the corannulene molecule and $F_E$ is the radiation field in the Red Rectangle \citep[adjusted from][]{Candian2015}. The energy $E$ is equal to the absorbed photon energy $E = h\nu$. The corannulene absorption cross-section ($\sigma_E$) is available from the PAH spectral database\footnote{\href{http://astrochemistry.ca.astro.it/database/}{http://astrochemistry.ca.astro.it/database/}} \citep{Malloci2007}. The Red Rectangle radiation field ($F_E$) is taken from the observational results of \cite{Vijh2005}. Although their data is not publicly available, the overall trend of the radiation field was obtained with the tool PlotDigitizer\footnote{\href{https://plotdigitizer.com/}{https://plotdigitizer.com/}}.

The internal energy distribution was finally calculated by normalising the absorption rate by the absorption rate integrated up to the Lyman limit of 13.6 eV:
\begin{equation}
    n(E) = \frac{R_{abs}(E)}{R_{abs}}
\end{equation}
.

\section{Overview of hydrocarbon abundances}
\label{dust_size_distribution}
\textbf{Table \ref{tab:abundances}} compares literature values of detected large hydrocarbon molecules with the upper limit for corannulene derived in this work. 

\begin{table*}
\centering
\begin{tabular}{lllll}
\textbf{Molecule} & \textbf{Environment} & \textbf{Abundance over hydrogen} & \textbf{Reported abundance} & \textbf{Source} \\ \hline
C$_{20}$H$_{10}$ & Post-AGB/PN (Red Rectangle Nebula) & < 5 $\times$ 10$^{-13}$ & - & This work \\
C$_{70}$ & Post-AGB/PN (Tc 1) & $7.3\times10^{-8}$ & 1.5\% of carbon & \citet{Cami2010} \\
C$_{60}$ & Post-AGB/PN (Tc 1) & $8.8\times10^{-8}$ & 1.5\% of carbon & \citet{Cami2010} \\
C$_{60}$ & Reflection Nebula (NGC 2073) & $7.9\times10^{-12} - 9.6\times10^{-10}$ & $ 1.4\times 10^{-4} - 1.7 \times 10^{-2}$\% of carbon  & \citet{Bern2011}  \\
C$_{42}$H$_{18}$ & Diffuse ISM & $< 1\times10^{-9}$ & $<2 \times 10^{-4}$ of carbon  & \citet{Kokkin2008}  \\
C$_{16}$H$_{10}$ & Cold Molecular Cloud (TMC-1) & $\sim1-2 \times 10^{-9}$ & $\sim3 \times 10^{-9}$ over H$_2$ & \citet{Wenzel2024} 
\end{tabular}
\caption{\textbf{An overview of reported large hydrocarbon abundances in various environments, including the corannulene upper limit derived in this work. The abundance estimates over carbon are converted to estimates over hydrogen, using a C/H ratio of $3.39\times 10^{-4}$ \citep{Zuo2021}.}}
\label{tab:abundances}
\end{table*}

%%%%%%%%%%%%%%%%%%%%%%%%%%%%%%%%%%%%%%%%%%%%%%%%%%

% Don't change these lines
\bsp	% typesetting comment
\label{lastpage}
\end{document}